# Long-range correlations
# in the statistical theory of critical fluid

## V.N. Bondarev[a)]

*Department of Theoretical Physics, Odessa I.I. Mechnikov National University,
2 Dvoryanskaya St., 65026 Odessa, Ukraine*

**Abstract**

Using the approach formulated in the previous papers of the author, a consistent procedure is developed for calculating non-classical asymptotic power terms in the total and the direct correlation functions of a "critical" fluid. Analyzing the Ornstein-Zernike equation allows us to find, for the first time, the values of transcendental exponents 1.73494 and 2.26989 which determine the asymptotic terms next to the leading one in the total correlation function. It is shown that already the simplest approximation based on only two asymptotic terms leads to the correlation functions, which are quantitatively close to the corresponding ones of the Lennard-Jones fluid (argon) in the near-critical state. The obtained results open a way for consistent theoretical interpretation of the experimental data on the critical characteristics of real substances. Both the theoretical arguments and analysis of published data on the experimentally measured critical exponents of real fluids lead to the conclusion that the known assumption of the sameness of the critical characteristics of the Ising model and the fluid in the vicinity of critical point (the "universality hypothesis") should be questioned.

## 1. Introduction

The total pair correlation function (PCF) $h(r)$ and the direct PCF $c(r)$ play fundamental role in the determination of the thermodynamic characteristics of statistical media [1–3]. Therefore, the search for approaches that allow one to calculate the transformations of PCFs in wide ranges of temperature and density, including the regions of phase transitions, is an actual problem. The effects in the vicinity of the gas-liquid critical point where the singular character of PCFs becomes apparent extremely brightly are of special interest [1–3]. However, the conventional approaches (such as the hyper-netted-chain and the Percus-Yevick approximations [2,3]) based on the integral equations of the statistical theory of liquids lead, in fact, to pure classical (by van der Waals [4]) values of the critical exponents or to values noticeably differing from those observed experimentally (see review [5]).

---

[a)] *E-mail address:* bondvic@onu.edu.ua

Numerous attempts to obtain non-classical critical exponents from the equations of modern theory of fluids were unsuccessful for a long time. Comparative analysis of the experimentally observed critical singularities of the thermodynamic functions of different media led in due time to the formulation of the so-called "universality hypothesis" (about it see, e.g., [6]). According to this hypothesis, any fluid in the near-critical state belongs to the same universality class $O(1)$ as the three-dimensional (3D) Ising magnet. Hence, it followed that to find non-classical critical exponents for fluid it is enough to study the critical characteristics of the 3D Ising (lattice gas) model by its numerical solution [5,7]. On the other hand, the renormalization group and $\varepsilon$-expansion methods [5,7,8] up to the recent past were the only analytical ones for finding the non-classical critical exponents which are compatible with the experimental data for fluids (see, e.g., [6,9]).

This, however, does not remove the question of whether it is possible to build a *non-classical* theory of the critical point *directly* from the fundamental equations of the theory of fluids. In paper [10] (and further in [11]), based on the formally closed equations for $h(r)$ and $c(r)$, the author has proposed an approach allowed to give the independent analytical calculation of the non-classical critical exponents of one-component non-Coulomb fluid. The exponents [10] have the form of rational fractions satisfying all known scaling relations, including the hyper-scaling relation. The values of the exponents [10,11] (except the exponent $\eta$, see below) are in rather good agreement with the corresponding values found by the numerical solution of the 3D Ising model [5–7]. Besides, the approach [10] was applied in [11] to the calculation of the next to the leading terms in the expressions for the correlation lengths (the so-called "correction-to-scaling" terms, following the terminology of [12]). Good quantitative agreement of the theory [10,11] and experiment opened a way to accurate processing of the experimental data on the critical properties of real substances (see [13–15]).

It is significant that the theory [10,11] allows further development and deepening in a direction which, in fact, has not been investigated until now. The point is that in addition to the mentioned "correction-to-scaling" terms, there is another source of the correction terms in the singular parts of the thermodynamic functions of the "critical" fluid. Really, let us represent, as usually (see, e.g., [4]), the total PCF near the critical point as follows:

$$h(r) = h_c(r) e^{-r/\xi}. \qquad (1)$$

Here, $\xi$ is the mentioned correlation length which tends to infinity when approaching the critical point, while the asymptotic ($r \to \infty$) behavior of the 3D "critical" total PCF $h_c(r)$ is determined by expression[1]

$$h_c(r) \sim r^{-(1+\eta)}, \qquad (2)$$

where, according to [10], $\eta = 1/5$. It is seen from (1) and (2) that next to the leading critical singularities of the thermodynamic quantities of the fluid will be determined not only by the "correction-to-scaling" terms in $\xi$ but also by next to the leading power-type asymptotic terms in $h_c(r)$. It is important that the theory [10,11] really allows us to find such terms in $h_c(r)$ and to calculate corresponding critical exponents, that could hardly be done by the renormalization group and $\varepsilon$-expansion methods [5,7,8].

---

[1] Here and below, considering the situation at the critical point itself, we mark the corresponding values by the subscript "$c$".

Thus, the purpose of the present paper is to go beyond the limits of the approximation in which the "critical" PCF $h_c(r)$ of real fluid is described by the only exponent $\eta$. In Section 2 (and also in Appendices A and B), being based on the fundamental equations for PCFs, we develop an approach which allows us to obtain, for the first time, the explicit form of the power-type asymptotic terms, which follow the leading one in $h_c(r)$. Using these results, as well as exact conditions on the "critical" PCFs, in Section 3 we find self-consistent expressions for these PCFs and demonstrate the possibility of applying them to interpreting (even at a quantitative level) experimental data for real fluid (argon) near the critical point. Further (Section 4), we argue that the approach developed in this paper (and in [10,11,13–15]) allows us to describe experimental data on the critical features of the thermodynamic functions of real fluids more accurately than the 3D Ising model. This means that the mentioned "universality hypothesis", which relates the "critical" fluid and the lattice gas (the Ising model) to one and the same universality class, hardly takes place. At last, in Section 4, we formulate the conclusions from the obtained theoretical results and note that their comprehensive experimental verification should be of fundamental interest.

## 2. Basic equations and asymptotic expansions of PCFs

The theory [10] is based on two formally exact equations of the statistical theory of liquids closed at the level of $h(r)$ and $c(r)$ PCFs. One of these equations obtained for the first time in [16,17] has the form:

$$\ln[h(r)+1] = h(r) - c(r) + B(r) - \frac{W(r)}{T}, \quad (3)$$

whereas the second is the well-known Ornstein–Zernike equation [1–3]:

$$h(r) - c(r) = \rho \int d\mathbf{r}' c(r') h(|\mathbf{r}'-\mathbf{r}|). \quad (4)$$

All information concerning many-particle correlations is contained in the so-called bridge function $B(r)$ represented by a complicated diagrammatic series in the form of integrals from the products of the total PCFs (see [16,17]). Equations (3) and (4) determine the equilibrium statistical characteristics of a one-component fluid with the atom number density $\rho$ and the temperature $T$. Having in mind non-Coulomb media, for $W(r)$, the energy of pair interaction between atoms separated by a distance $r$, one assumes the Lennard-Jones potential

$$W(r) = 4w_0\left[\left(\frac{D}{r}\right)^{12} - \left(\frac{D}{r}\right)^{6}\right], \quad (5)$$

where $w_0$ and $D$ are characteristic constants (for argon, $w_0 = 0.0104$ eV and $D = 3.40$ Å [18]).

A *key point* of the theory [10] is the representation of the asymptotic (at $r \to \infty$) behavior of $B(r)$ in the form of a local expansion:

$$B(r) = a_2(T,\rho)h^2(r) + a_3(T,\rho)h^3(r) + a_4(T,\rho)h^4(r) + ... \qquad (6)$$

with coefficients $a_j(T,\rho)$ ($j$=2, 3, 4,...) which depend on the thermodynamic state of the medium. In particular cases, the representation (6) leads to the well-known approximations, e.g., the hyper-netted-chain approximation [2,3] at $a_j(T,\rho) \equiv 0$ or the mean-spherical approximation [5,19] at $a_j(T,\rho) \equiv (-1)^{j+1}/j$.

Refusal of such harsh conditions on the coefficients in (6) leads to the theory [10,11] in which a *non-classical* asymptotic behavior of the "critical" PCFs follows from the fundamental equations (3) and (4). In this case, the two critical values, $T_c$ and $\rho_c$, must satisfy the two conditions [10]:

$$a_2(T_c,\rho_c) = -\frac{1}{2}, \quad a_3(T_c,\rho_c) = \frac{1}{3}, \qquad (7)$$

which lead to the vanishing of the terms $\sim h_c^2(r)$ and $\sim h_c^3(r)$ in the asymptotic limit of equation (3). Then the first non-vanishing asymptotic term on $h_c(r)$ in equation (3) is $(a_{4c}+1/4)h_c^4(r)$, where we have designated $a_{4c} \equiv a_4(T_c,\rho_c)$. Thus, the asymptotic form of equation (3) at the critical conditions is as follows:

$$c_c(r) = \left(a_{4c} + \frac{1}{4}\right)h_c^4(r) + \left(a_{5c} - \frac{1}{5}\right)h_c^5(r) + ... - \frac{W(r)}{T_c}, \qquad (8)$$

where we have written the first non-vanishing terms of the expansion in powers of $h_c(r)$. We shall see (Appendices A and B) that for our purposes it is sufficient to consider only the term $\sim h_c^4(r)$ in (8).

Further, we shall use the Fourier representation of the Ornstein–Zernike equation (4) (see, e.g., [3])

$$\tilde{h}(k) = \frac{\tilde{c}(k)}{1 - \rho\tilde{c}(k)}, \qquad (9)$$

where the Fourier-component of the total PCF of the 3D fluid is

$$\tilde{h}(k) = \int d\mathbf{r}\, h(r) e^{-i\mathbf{k}\mathbf{r}} = \frac{4\pi}{k} \int_0^\infty dr\, r \sin(kr) h(r) \qquad (10)$$

and we can write similar expression for $\tilde{c}(k)$. Also, we shall need the equation

$$T\kappa_T = \frac{1}{\rho} + \tilde{h}(0) \tag{11}$$

for the isothermal compressibility $\kappa_T = \rho^{-1}(\partial\rho/\partial p)_T$ (see, e.g., [3]).

Now, let us represent the leading asymptotic term in the "critical" direct PCF in the form analogous to (2):

$$c_c(r) \sim r^{-m}, \tag{12}$$

with some exponent $m$. Then using the form (8), analyzing equation (9) at the critical point itself and taking into account that $\tilde{c}_c(0) = 1/\rho_c$ (see, e.g., [3]) one can arrive at the values $\eta = 1/5$ and $m = 24/5 < 6$ for real 3D fluid [10].

Below we shall show that two more asymptotic terms in $h_c(r)$ will be also determined just by the term $\sim h_c^4$ in (8), i.e. when finding the corresponding critical exponents, potential $W(r)$ of the form (5) should not at all be taken into account. Meanwhile, the critical values $T_c$ and $\rho_c$ will be determined just by the parameters of this potential.[2]

The dependences (2) and (12) represent the leading asymptotic singular terms in the PCFs of the "critical" fluid. In this Section and in the Appendices A and B we develop a procedure for investigating the analytic properties of the Ornstein–Zernike equation (9) and find also the further singular terms of the "critical" PCFs. It is essential that in order to find the first two terms after the leading one, we should not go beyond the approach based on the account of the term $\sim h_c^4(r)$ in the "critical" direct PCF (8).

The asymptotic term $\sim r^{-n}$ in $h_c(r)$ with $n \equiv 1+\eta = 6/5$ [10] determines the leading divergent contribution into the isothermal compressibility (11) of the "critical" fluid. From (10) and (11) it follows that the terms in $h_c(r)$ with the asymptotic behavior $\sim r^{-n'}$ under the condition $6/5 < n' < 3$ will also lead to the divergent contributions into $\kappa_T$.

So, we represent the "critical" total PCF at $r \to \infty$ in the form:

$$h_c(r) = \frac{A_\infty}{r^{6/5}} + \frac{A'_\infty}{r^{n'}} \tag{13}$$

with some constants $A_\infty$ and $A'_\infty$; as for the exponent $n'$, we shall consider it as the smallest permissible value from the "dangerous" interval $6/5 < n' < 3$. Substituting (13) into (8) we find the asymptotic form of the "critical" direct PCF up to the main term containing the exponent $n'$:

$$c_c(r) = \left(a_{4c} + \frac{1}{4}\right)\left(\frac{A_\infty^4}{r^{24/5}} + 4\frac{A_\infty^3 A'_\infty}{r^{18/5+n'}}\right). \tag{14}$$

---

[2] In paper [15] the author has demonstrated how these critical values can be restored in the framework of the non-classical theory of criticality when using the known numerical calculations of the virial coefficients for the Lennard-Jones fluid; see also [20].

It is important to note that if $n' < 12/5$, the asymptotic term $\sim r^{-18/5-n'}$ in (14) decreases more slowly than the last two terms in the expression (8). Therefore, below we will find the exponent $n'$ in the range $6/5 < n' < 12/5$.

Somewhat bulky calculations in Appendix A lead to equality:

$$n' = 1.73494.... \qquad (15)$$

*Thus, in the framework of the theory* [10,11] *we succeeded in finding the numerical value of the exponent $n'$, determining the asymptotic behavior of the next to the leading term in the singular part of the "critical" PCF* (13). The found exponent (15) is transcendental, unlike the leading exponent $n \equiv 1 + \eta = 6/5$ [10]. Since $n' < 12/5$ the asymptotic term $\sim r^{-18/5-n'}$ in (14) decreases slower than the last two terms in (8) and therefore is really the most important after the leading term $\sim r^{-24/5}$.[3]

The terms written in (13) do not exhaust the singular ones in $h_c(r)$ that lead to the divergent contributions into the isothermal compressibility of the "critical" fluid. Notable interest is the calculation of one more power term decreasing "faster" than those written in (13). Adding a term with exponent $n'' > n'$ into the "critical" total PCF we can present its asymptotic form as follows:

$$h_c(r) = \frac{A_\infty}{r^{6/5}} + \frac{A'_\infty}{r^{n'}} + \frac{A''_\infty}{r^{n''}}, \quad r \to \infty, \qquad (16)$$

where $A''_\infty$ is a new constant. The calculation of the exponent $n''$ given in the Appendix B leads to the value:

$$n'' = 2n' - \frac{6}{5} = 2.26989... \qquad (17)$$

together with definite relation (B9) allowing us to express the constant $A''_\infty > 0$ through the constants $A_\infty > 0$ and $A'_\infty$.

The value of $n''$ is still less than $12/5$, so the corresponding asymptotic contribution into $c_c(r)$ will decrease "slower" than the last two terms in (8). Note, that the exponent $n''$ as well as $n'$ is transcendental. One can make sure that the terms in (16) with the found exponents $n = 6/5$ [10], $n'$ (see (15)), and $n''$ (see (17)) exhaust all the power terms, the origin of which is due to the term $\sim h_c^4(r)$ in (8).

## 3. The self-consistent PCFs of the Lennard-Jones fluid at criticality: an example

Now we apply the obtained results to construct approximate PCFs of the Lennard-Jones fluid *directly at the critical point*. To do this, we use the asymptotic expression (6) for the "critical" bridge function, as well as the critical conditions (7), and approximately represent $B_c(r)$ over the entire interval $0 < r < \infty$ in the form:

$$B_c(r) = -\frac{1}{2}h_c^2(r) + \frac{1}{3}h_c^3(r) + a_{4c}h_c^4(r). \tag{18}$$

---

[3] It is curious to note, that the term $\sim h_c^5(r)$ in equation (8) at $\eta = 1/5$ decreases in the asymptotic region like $r^{-6}$, i.e. as the Lennard-Jones potential (5).

We shall also use the so-called cavity PCF $y(r)$, coupled with $h(r)$ by the relation (see, e.g., [21]):

$$1 + h(r) = [1 + f(r)]y(r). \tag{19}$$

Here $f(r) \equiv e^{-W(r)/T} - 1$ is the well-known Mayer function (see, e.g., [3]), and also $y(r \to \infty) \to 1$. In contrast to the Mayer function defined by the interatomic potential (5), the function $v_c(r) \equiv y_c(r) - 1$, like $h_c(r)$, is long-range at the critical point, i.e. $v_c(r) \sim r^{-6/5}$ when $r \to \infty$. Therefore, in order to find an approximate form of the function $h_c(r)$ that has a correct behavior at $r \to 0$ and $r \to \infty$, it is enough to specify the correct asymptotic behavior of the function $v_c(r)$.

Having in mind expression (13) and taking into account the value (15) for the next to the leading exponent of the asymptotic decay of $h_c(r)$, we represent the simplest approximate expression for the function $v_c(r)$ at $0 < r < \infty$ as follows:

$$v_c(r) = \frac{A_\infty}{(r^2 + r_0^2)^{3/5}} + \frac{A'_\infty}{(r^2 + r_0^2)^{n'/2}}. \tag{20}$$

The constants $A_\infty > 0$, $A'_\infty$, and $r_0^2$ entering in (20) will be determined from the integral conditions to which the correlation functions must satisfy.

Two of these conditions were formulated in [11], starting from the requirement that the first three density derivatives of the fluid pressure at the critical point vanish (and also the finiteness of the fourth derivative required in the theory [10,11,15]). In the above notation, these two conditions have the form:

$$\int_0^\infty dr\, r^4 \frac{dW(r)}{dr}[1 + f_c(r)]v_c(r) = 0, \tag{21}$$

$$\int_0^\infty dr\, r^5 \frac{dW(r)}{dr}[1 + f_c(r)]v_c(r) = 0. \tag{22}$$

As the third condition, it is natural to choose the requirement of the divergence of the isothermal compressibility (11) at the critical point of the fluid. This requirement, taking into account (9), reads:

$$4\pi\rho_c \int_0^\infty dr\, r^2 c_c(r) = 1. \tag{23}$$

With the help of (3), (6), and (7), one can represent the approximate PCF $c_c(r)$ entering the critical condition (23) in the form:

$$c_c(r) = -\ln[1+v_c(r)] + h_c(r) - \frac{1}{2}h_c^2(r) + \frac{1}{3}h_c^3(r) + a_{4c}h_c^4(r). \qquad (24)$$

In this case, the definition (19), as well as the relation (A14) should be taken into account. In what follows we shall use the dimensionless quantities: $\rho^* = \rho D^3$, $T^* = T/w_0$, $A_\infty^* \equiv A_\infty/D^{6/5} > 0$, $A_\infty'^* \equiv A_\infty'/D^{n'}$ and the values of the critical parameters $\rho_c^* = 0.3449$, $T_c^* = 1.3303$ found in [15]. Substituting the values of integrals (A4) into (A13), we obtain for $\rho_c^* = 0.3449$ the following relation between $a_{4c}$ and the required constant $A_\infty^* > 0$:

$$a_{4c} = \left(\frac{0.5364}{A_\infty^*}\right)^5 - \frac{1}{4}. \qquad (25)$$

It is clear from the structure of expression (24) that the desired function (20) must be such that the inequality $v_c(r) > -1$ (following from the general inequality $v(r) > -1$, see (3)) would be satisfied in the whole domain $0 < r < \infty$. Below we will use the dimensionless variable $\hat{r} \equiv r/D$ and the dimensionless constant $\hat{r}_0 \equiv r_0/D$. Then equations (21) and (22) can be written as a system of linear homogeneous equations with respect to the unknown quantities $A_\infty^* > 0$ and $A_\infty'^*$:

$$A_\infty^* g_1(\hat{r}_0) + A_\infty'^* g_1'(\hat{r}_0) = 0, \qquad (26)$$

$$A_\infty^* g_2(\hat{r}_0) + A_\infty'^* g_2'(\hat{r}_0) = 0. \qquad (27)$$

The coefficients in these equations depend on $\hat{r}_0$ and have the form:

$$g_1(\hat{r}_0) = \int_0^\infty d\hat{r} \frac{1}{\hat{r}^3}\left(1 - \frac{2}{\hat{r}^6}\right)\frac{1+f_c(\hat{r})}{(\hat{r}^2+\hat{r}_0^2)^{3/5}}, \quad g_1'(\hat{r}_0) = \int_0^\infty d\hat{r} \frac{1}{\hat{r}^3}\left(1 - \frac{2}{\hat{r}^6}\right)\frac{1+f_c(\hat{r})}{(\hat{r}^2+\hat{r}_0^2)^{n'/2}}, \qquad (28)$$

$$g_2(\hat{r}_0) = \int_0^\infty d\hat{r} \frac{1}{\hat{r}^2}\left(1 - \frac{2}{\hat{r}^6}\right)\frac{1+f_c(\hat{r})}{(\hat{r}^2+\hat{r}_0^2)^{3/5}}, \quad g_2'(\hat{r}_0) = \int_0^\infty d\hat{r} \frac{1}{\hat{r}^2}\left(1 - \frac{2}{\hat{r}^6}\right)\frac{1+f_c(\hat{r})}{(\hat{r}^2+\hat{r}_0^2)^{n'/2}}. \qquad (29)$$

Equating the determinant of the system (26) and (27) to zero, we obtain the condition for the nontrivial solvability of this system:

$$g_1(\hat{r}_0)g_2'(\hat{r}_0) - g_2(\hat{r}_0)g_1'(\hat{r}_0) = 0. \qquad (30)$$

This condition is the equation for finding $\hat{r}_0$. Knowing $\hat{r}_0$ and using equation (26), one can obtain a relation $A_\infty'^* = -A_\infty^* g_1(\hat{r}_0)/g_1'(\hat{r}_0)$ between the remaining two quantities. If we substitute this relation and the found value of $\hat{r}_0$ into expressions (20) and (24), take into account (25) and use the requirement (23), then as a result we obtain an equation for determining the constant $A_\infty^* > 0$, after which we also find the constant $A_\infty'^*$. Within this scheme, using a simple numerical procedure, the following values of the dimensionless constants were found: $\hat{r}_0 = 1.5247$, $A_\infty^* = 0.6627$, $A_\infty'^* = -1.1732$. In addition, based on representations (18) and (24) with the found value $a_{4c} = (0.5364/A_\infty^*)^5 - 1/4 = 0.09746$, one can construct the desired approximate expression for the "critical" bridge function $B_c(r)$.

The obtained results are enough to calculate the "critical" PCFs $h_c(r)$ and $c_c(r)$ in the framework of the developed approximation. Unfortunately, there is no data in the literature concerning the form of PCFs for fluids at criticality, which is associated with large (and, in fact, insurmountable) difficulties in carrying out real experiments or computer simulations directly at the critical point. So, most informative seems the comparison of the obtained theoretical results with the experimental data for real fluids in the near-critical region. Apparently, the old papers [22,23] dedicated to the study of x-ray scattering on argon are most suitable for this (see also [24]). The PCFs of argon $h(r)$ [22] and $c(r)$ [23] were recovered by the experimental data processing both on isochore at mass density 0.536 g/cm$^3$ (close to the critical value, 0.531 g/cm$^3$ [25]), and on isotherm at temperature $T \approx 148$ K (close to the critical value, 150.65 K [25]).

In Fig. 1, the solid (red) line is built by the results of our calculation of $h_c(\hat{r})$. By the dashed lines (between which the measurement error interval is enclosed) we show the values of $h(\hat{r})$ built using the experimental data [22] for argon at mass density 0.536 g/cm$^3$ and temperature $T \approx 153$ K close to the corresponding critical values.

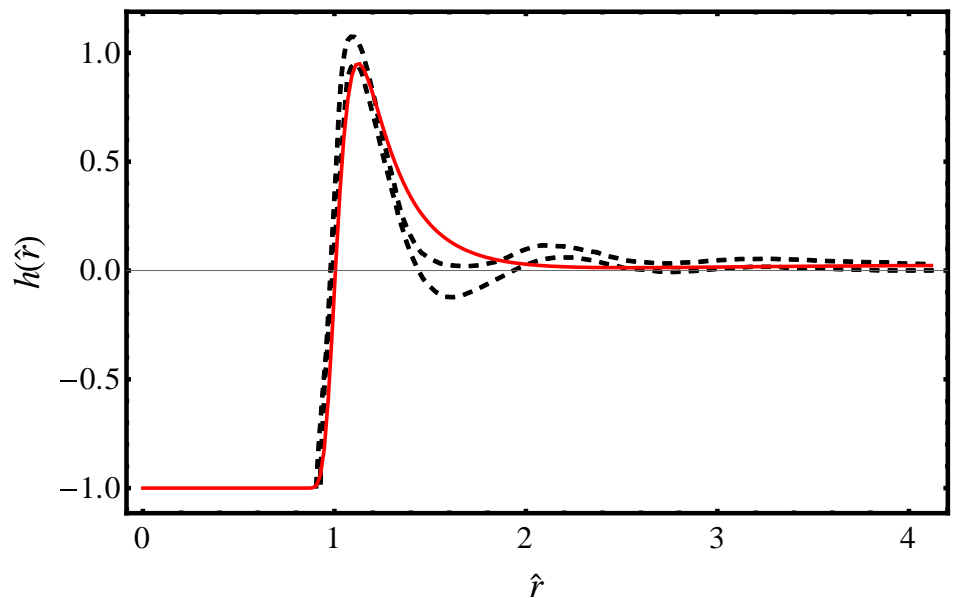

Fig. 1. The total PCF $h_c(\hat{r})$ calculated by our theory at criticality (solid red line) in comparison with the total PCF $h(\hat{r})$ extracted from the experimental data [22] for argon at near-critical conditions (dashed lines, between which the measurement error interval is enclosed).

Fig. 2, in the same notations and at the same values of the mass density and temperature, demonstrates the results of our calculation of $c_c(\hat{r})$ (solid red line) in comparison with the experimental (with taking account of measurement error) dependence of $c(\hat{r})$ for argon [23]. It can be seen that the general course of the theoretical dependences $h_c(\hat{r})$ and $c_c(\hat{r})$ is in satisfactory agreement with the experimental data given [22,23], although smaller details (the "fine structure" of $h_c(\hat{r})$) are lost.

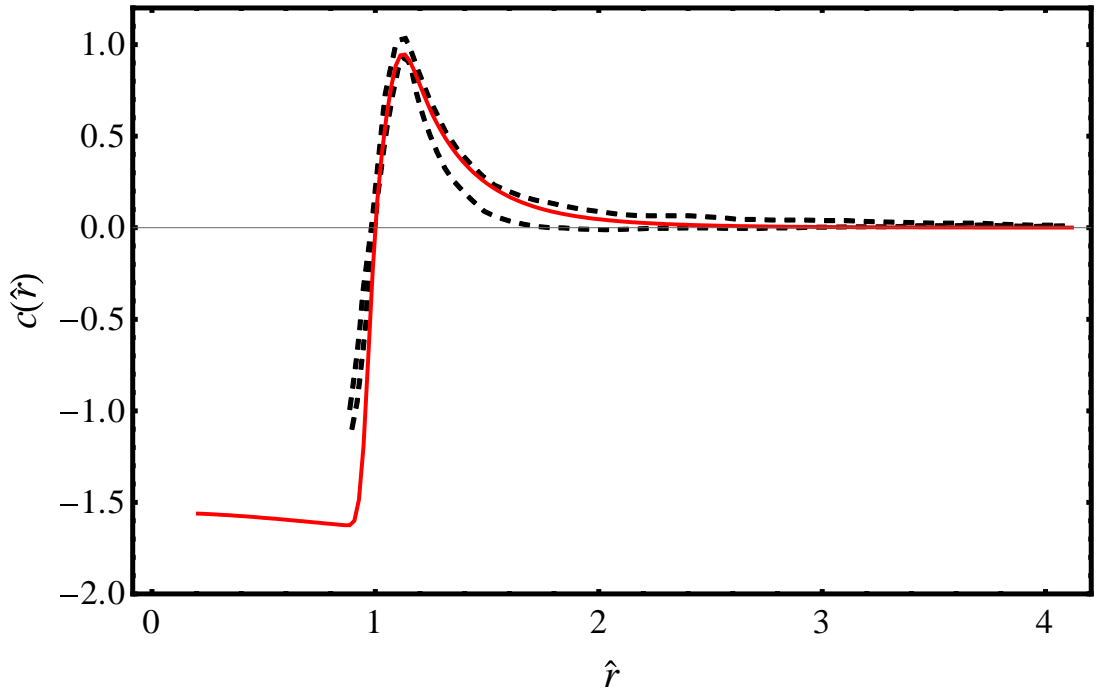

Fig. 2. The direct PCF $c_c(\hat{r})$ calculated by our theory at criticality (solid red line) in comparison with the direct PCF $c(\hat{r})$ extracted from the experimental data [23] for argon at near-critical conditions (dashed lines, between which the measurement error interval is enclosed).

Thus, already within the framework of the simplest approximation, our theory makes it possible even at a quantitative level to grasp the most characteristic features (including the locations and heights of the maxima) of the "critical" PCFs of the Lennard-Jones fluid. Herewith, we used only the theoretically calculated values of the two critical exponents (the leading [10] and the next to the leading (15)), describing the asymptotic decay of $h_c(\hat{r})$, and the exact integral conditions for PCFs.

As for the "critical" pressure $p_c$, then in the adopted rough approximation it turns out to be negative: $p_c^* \equiv p_c D^3 / w_0 \approx -0.059$ (whereas the calculation [15] using the virial expansion gives positive value: $p_c^* \approx 0.136$). The origin of the formally negative $p_c^*$ sign becomes clear given the following circumstance. The fact is that the oversimplified approximation based on (20) leads to an overestimated contribution of the long-range correlations in $p_c^*$, since at large enough distances the interatomic interaction has the character of attraction (cf. [15], where the question of the negative sign of the singular contribution to the "critical" pressure is discussed).

It is important, that the results obtained in this paper allow us to refine the theoretical description of the "critical" PCFs by using a term with one more critical exponent $n''$ (see (17)). It can be expected that due to the appearance of an additional term in the expression for $v_c(r)$ it will be possible to reproduce in more detail the experimentally observed features of the structure of PCFs at criticality. The same applies to the "critical" pressure, the calculation of which in a more perfect approximation can lead to a value consistent with that found, for example, in [15].

It is useful to mention the recent work [26], in which the results of numerical simulation of the functions $1+h(\hat{r})$, $y(\hat{r})$, and $B(\hat{r})$ for the Lennard-Jones fluid have been demonstrated at temperatures close to $T_c^*$ but at densities $\rho^* = 0.177$ (much lower than $\rho_c^*$) and $\rho^* = 0.462$ (much higher than $\rho_c^*$). Therefore, one can only perform a qualitative comparison of our results with the data of [26]. For example, the heights of the main peak of the function $h(\hat{r})$ at both indicated values of the density [26] were close to each other and found to be in numerical agreement with the height of the corresponding peak for our PCF $h_c(\hat{r})$ represented in Fig. 1. However, the structure of $h(\hat{r})$ at $\hat{r} > 1.5$, according to [26], is more pronounced than that of our $h_c(\hat{r})$. As for the "critical" bridge function $B_c(r)$, the calculations by Eq. (18) show that for all $r$ it remains negative and is in qualitative agreement with the bridge functions from [26]. Note, that although the "critical" bridge function (18) is a long-range one, its asymptotic decay follows the law $-r^{-12/5}$, i.e. turns out to be much "faster" than the decay of $h_c(r)$.

## 4. Discussion of the results. Conclusions

Here, it is appropriate to touch upon the question of how adequately it is possible to describe the critical characteristics of real fluids using existing theoretical approaches. To date, extensive material has been accumulated in the literature on the critical properties of pure liquids and liquid mixtures. Beginning in the 70s of the 20th century, to interpret the experimentally observed critical features of the thermodynamic quantities of real fluids, the results of the numerical analysis of 3D lattice gas model (isomorphic to the Ising magnet) became to be actively used. However, the "universality hypothesis", according to which the "critical" fluid can be described by the 3D Ising model, remained unproved. Below, based on known experimental data, we will argue that real fluids in the vicinity of the liquid-gas critical point should be placed into a special universality class that does not coincide with the universality class of the 3D Ising model. At the same time, as an alternative to the lattice gas model, we will use the results of our approach [10,11], based on the fundamental equations of the theory of liquids.

Although the critical exponents [10] turned out to be rather close to the corresponding exponents of the 3D Ising model there are numerical differences between them (especially for the exponent $\eta$, see above).[4] This result can be considered as an additive argument in favor of the conclusion that real "critical" fluid and the 3D Ising magnet can hardly belong to one and the same universality class.[5]

For today, one can point out quite a few experimental evidences that the critical exponents of real fluids in the asymptotic region (immediately adjacent to the critical point) are not in precise accordance with those found by the numerical solution of the 3D Ising model. Furthermore, attempts to fit the experimental data using formulas which contain, along with the leading Ising term, a few more Wegner [12] "correction-to-scaling" terms often lead to less accurate description than approximations by formulas containing sometimes the only one term of non-Ising type. In addition, formulas with non-Ising exponents are capable to give the quantitative description of experiments in a much wider area of thermodynamic variables than analogous formulas with the 3D Ising exponents. Let us give some of the most reliable evidences in favor of the non-Ising interpretation of the experiments on the critical properties of real fluids.[6]

a) Exponent $\alpha$ characterizes the temperature singularity of the heat capacity both along the critical isochore and along the liquid-gas coexistence curve: $C_V \sim |t|^{-\alpha}$, $t \equiv (T/T_c - 1)$. The 3D Ising model calculations give $\alpha = 0.109$ [6]; the theory [10] leads to the value $\alpha = 1/8$.

In paper [28], an attempt of numerical processing of the most accurate (in a microgravity environment during the German Spacelab Mission D-2) $C_V$ measurements in the near-critical region of $SF_6$ had been made.[7] The asymptotic region was estimated in [28] as $|t| < 1.6 \times 10^{-4}$, whereas the whole temperature interval of measurements was $3 \times 10^{-6} < |t| < 10^{-2}$. Later, in [29], the experimental data [28] were reanalyzed and described with high accuracy up to $|t| \cong 10^{-2}$ using the *simple* power law (see also [6]). The conclusion of [29] reads: "These heat capacity measurements were found to be inconsistent with renormalization group predictions". Furthermore, according to [6], "all data" of [28] "were actually within the asymptotic region". Serious arguments were presented in [30] in favor of the existence of the so-called "Yang-Yang anomaly" (the divergence of the second temperature derivatives of both the pressure and the chemical potential in the expression for $C_V$ [31]) in the vicinity of $^3$He critical point. Then, according to [30], "…a revision of the conventional scaling theory for fluids based on the analogy between fluids and lattice-gas model will be required".[8]

It is also useful to mention the article [34], in which the measured value $\alpha = 0.124 \pm 0.006$ was given for the heat capacity exponent of the liquid-liquid mixture nitrobenzene-dodecane near its critical consolute point.

---

[4] The differences between the two-dimensional (2D) critical exponents calculated using theory [10], on the one hand, and the corresponding Onsager's exponents for the 2D Ising model [4], on the other hand, are more striking than for the 3D case. A detailed discussion of this question we leave for future.

[5] In this regard, it is worth mentioning the paper [27] containing the next statement: "However, nothing is proven about liquid-gas criticality from the universality conjecture".

[6] An extensive collection of measured critical exponents for pure liquids and mixtures is given in books [9,25].

[7] Unfortunately, the outlined in [6] program on microgravity study of other thermodynamic characteristics, such as the coexistence curve, the isothermal compressibility, etc., apparently, still has not been implemented to "critical" fluids.

[8] Moreover, the "Yang-Yang anomaly" was detected repeatedly in other systems (see, for example, papers [32,33]).

In addition, in work [35] a path is outlined, allowing to establish a trend to the divergence of the isochoric heat capacity of a fluid near the liquid-gas critical point based on the results of molecular dynamics simulation, including also the effects of three-body interaction.

b) Exponent $\beta$ at $t < 0$ characterizes the critical singularity of the liquid-gas coexistence curve: $|\rho_{l,g} - \rho_c| \sim |t|^{\beta}$, where $\rho_l$ and $\rho_g$ are the fluid densities at the liquid and gas branches, respectively. The calculations by the 3D Ising model [6] give $\beta \approx 0.325$; according to the theory [10] $\beta = 3/8$.

The experimental coexistence curves of Ne, HD, $N_2$, and $CH_4$ were analyzed [36] in the interval $1.3 \times 10^{-4} < |t| < 2 \times 10^{-2}$ using the *single*-term fitting with $\beta = 0.355$. As the result, the conclusion [36] reads: "We draw attention to the disadvantages of using the extended scaling equations with the Ising exponent $\beta$ for describing coexistence curves of simple one-component fluids". For a number of "critical" fluids the exponent $\beta$ is at the level of $\beta = 0.34$–0.36 [9,25, 37–44] reaching for Heptane even the value $\beta = 0.385 \pm 0.016$ [45] far exceeding the Ising value. One should also mention paper [46] devoted to molecular dynamics simulation of the liquid-gas criticality. It is essential, that without the use of the so-called "finite size scaling" approach, the *non-Ising* exponent $\beta = 0.35$–0.36 for a fluid with the Lennard-Jones interaction was obtained in [46].

In our papers [13,14], in addition, it has been demonstrated the possibility of accurate processing of the experimental data for a number of substances just with the exponent $\beta = 3/8$ found in [10].

c) Exponent $\nu$ determines the divergence of the fluid correlation length on the critical isochore: $\xi \sim |t|^{-\nu}$. The value $\nu = 5/8$ in the theory [10] is very close to the corresponding value $\nu \approx 0.623$–0.63 for the 3D Ising model [5]. In paper [10] the author has shown that the light-scattering experiments near the critical point of $^3$He can be well described quantitatively by $\nu = 5/8$, at least when $|t| < 3 \times 10^{-3}$. The last value agrees on the order of magnitude with the above restriction on the asymptotic region near the critical point of real fluids.

d) Regarding the critical exponents $\gamma$ and $\delta$ which determine, respectively, the divergence of the isothermal compressibility $\kappa_T \sim |t|^{-\gamma}$ and the form of the critical isotherm $|p - p_c| \sim |\rho - \rho_c|^{\delta}$ in the pressure – density plane one can find significant discrepancies in the literature. Often, the experimental data processing (see, for example, [39,42,47]) led to a conclusion that these exponents differ from the corresponding values $\gamma \approx 1.24$ and $\delta \approx 4.8$ for the 3D Ising model [6], being closer to the values $\gamma = 9/8$ and $\delta = 4$ found in the theory [10].

Papers [48–52] reported on the measurements of non-Ising critical exponents for noble fluids: $\gamma \approx 1.10$–1.18, $\delta \approx 3.8$–4.21, as well as $\beta = 0.354$–0.361. It is interesting, that in the long-standing paper [53] dedicated to the study of critical properties of xenon, carbon dioxide, and hydrogen, it was found that "the critical isotherm is of the fourth degree", more precisely, $\delta \approx 4.1 \pm 0.2$.

e) Exponent $\eta$ determines, according to (2), the asymptotic (at large distances $r$) behavior of the total PCF of the 3D "critical" fluid. Remind that, according to the theory [10], exponent $\eta = 1/5$ is noticeably greater than the value $\eta \approx 0.035$ [5] found in the framework of the 3D Ising model.

The published experimental estimates for $\eta$ are in a fairly wide range – from $\eta \approx 0.05$ to $\eta = 0.15$ [39, 54–58], $\eta = 0.2 \pm 0.1$ (see the discussion of this question in [59]) and even higher [60]. Also it is useful to cite the conclusion of [55]: "*Our* *experiment clearly demonstrates that the ordinary lattice-gas model cannot be used to calculate the critical exponent η for a fluid*". Thus, the value $\eta = 1/5$ derived in [10] falls within the boundaries of the above experimental data.

So, *the analysis of the experimental data obtained by long-term studies of critical properties of real fluids, gives good reason to doubt the "universality hypothesis", which attributes the fluid near the liquid-gas critical point to the universality class O*(1) *of the 3D Ising model*. In fact, this means that "critical" fluids should be placed in a separate universality class, the singular properties of which were investigated in papers [10,11,13–15] (as well as in the present paper) using an approach based on fundamental equations of the modern theory of liquids.

Now, let us summarize the results obtained in the present paper. A regular procedure is developed which allows one calculating the next to the leading non-classical asymptotic terms in the "critical" PCFs $h_c(r)$ and $c_c(r)$ of real fluid. Based on the approach formulated in the author's papers [10,11], it is shown that not only the leading asymptotic term in $h_c(r)$ but also the following two terms giving the divergent contributions to the isothermal compressibility of fluid are determined just by the term $\sim h_c^4(r)$ in the asymptotic expansion of $c_c(r)$. Studying the analytical properties of the Ornstein-Zernike equation we, for the first time, succeeded in finding the transcendental values of the next to the leading critical exponents $n' = 1.73494...$ and $n'' = 2.26989...$ for $h_c(r)$.

As a useful application of the obtained results, the new term $\sim r^{-n'}$ (in addition to the term $\sim r^{-6/5}$ [10]) is taken into account to construct a simple approximate form of $h_c(r)$, and with it, $c_c(r)$. It is significant that for finding the characteristic parameters defining the form of $h_c(r)$ and $c_c(r)$, only strict integral conditions on the "critical" PCFs are used. It is shown that, already in the accepted simple approximation, the theory is capable even at a quantitative level to reproduce the characteristic features of PCFs (in particular, the positions and heights of the main maxima) of argon in the near-critical region. It is important that the theory allows for further improvement of the approximate description of the "critical" PCFs, in particular, by including the new term $\sim r^{-n''}$ into $h_c(r)$. Thus, the use of the results obtained in the present paper opens a way for the consistent theoretical interpretation of the experimentally observed peculiarities of the thermodynamic quantities of real fluids in the vicinity of the liquid-gas critical point.

The literature data analysis performed in the present paper leads to the conclusion that the experimentally measured values of the critical exponents for real fluids, usually, *do not coincide* with the corresponding values calculated in the framework of the 3D Ising model. Taking into account the arguments given, *it is necessary once again to consider the question about the sameness of the critical characteristics of the Ising magnet and of fluid near the liquid-gas critical point (the "universality hypothesis")*. Note that a definite answer to this question could be given by the "microgravity" measurements of the critical exponents *β, γ, δ* for fluids. Although up to the present time such measurements have been carried out only for the specific heat of $SF_6$ (exponent *α*), their results [27,28] clearly show the existence of the extended temperature asymptotic region in which "correction-to-scaling" terms are insignificant. The fact that using the critical exponents found in the papers [10,11] one achieves (see [13,14]) better agreement with the experiment than using the 3D Ising model exponents, *should be considered as an additive evidence against the*

*"universality hypothesis" in the context of real fluids criticality.* Therefore, the implementation of new precision experiments in the close vicinity of the liquid-gas critical point of real substances is of fundamental interest.

**Acknowledgements**

The author wishes to thank Dr. S.I. Kosenko and Dr. V.V. Zavalniuk for help in the numerical calculations.

**Appendix A. The calculation of critical exponent $n'$**

Equation (9) at the critical point, itself, can be rewritten as follows:

$$\tilde{h}_c(k) = \frac{\tilde{c}_c(0)\tilde{c}_c(k)}{\tilde{c}_c(0) - \tilde{c}_c(k)}. \tag{A1}$$

To analyze the behavior of $\tilde{h}_c(k)$ in the limit $k \to 0$, we represent $h_c(r)$ in the form:

$$h_c(r) = \frac{A_c(r)}{r^{6/5}} + \frac{A'_c(r)}{r^{n'}}, \tag{A2}$$

where we have introduced the functions $A_c(r)$ and $A'_c(r)$ which formally provide the proper behavior of $h_c(r)$ at any $r$; at that, according to (13), $A_c(r \to \infty) \to A_\infty$, $A'_c(r \to \infty) \to A'_\infty$.

Now, let us substitute (A2) into (10) and pass to new variable of integration $\zeta = kr$:

$$\tilde{h}_c(k) = \frac{4\pi}{k} \int_0^\infty dr\, r \sin(kr) h_c(r)$$
$$= 4\pi \int_0^\infty d\zeta\, \zeta \sin\zeta \left[ k^{-9/5} \frac{A_c(\zeta/k)}{\zeta^{6/5}} + k^{n'-3} \frac{A'_c(\zeta/k)}{\zeta^{n'}} \right]. \tag{A3}$$

At $k \to 0$, the value of the integral in (A3) will be determined by the behavior of the functions $A_c(\zeta/k)$ and $A'_c(\zeta/k)$ at $(\zeta/k) \to \infty$, i.e. by their asymptotic values $A_\infty$ and $A'_\infty$. Thus, the main contributions to $\tilde{h}_c(k)$ at $k \to 0$ in the interval $6/5 < n' < 12/5$ will be given by the expression:

$$\tilde{h}_c(k) = 4\pi A_\infty I_{6/5} k^{-9/5} + 4\pi A'_\infty I_{n'} k^{n'-3} \tag{A4}$$

which contains the positive integrals with the Euler's gamma-function [61]:

$$I_s \equiv \int_0^\infty d\zeta \frac{\sin \zeta}{\zeta^{s-1}} = \Gamma(2-s)\sin\left(\frac{\pi s}{2}\right), \quad 1 < s < 3. \tag{A5}$$

The analysis of expression (A4) allows us to claim that

$$A_\infty > 0. \tag{A6}$$

Really, the isothermal compressibility $\kappa_T$ (see (11)) must be positive for any real equilibrium system [4]. On the other hand, the singular part of $\kappa_T$ directly at the critical point can be considered as the limit $\tilde{h}_c(k \to 0)/T_c$ where the leading singularity will be determined by the first term in the right-hand side of expression (A4). As the result, condition (A6) follows.

Now, let us consider the function

$$c_c(r) = \left(a_{4c} + \frac{1}{4}\right)\left[\frac{A_c^4(r)}{r^{24/5}} + 4\frac{A_c^3(r)A_c^{'}(r)}{r^{18/5+n'}}\right] \tag{A7}$$

with the asymptotic behavior (14) and then let us study the difference

$$\tilde{c}_c(0) - \tilde{c}_c(k) = 4\pi\left(a_{4c} + \frac{1}{4}\right)\int_0^\infty dr\, r^2 \left[1 - \frac{\sin(kr)}{kr}\right]\left[\frac{A_c^4(r)}{r^{24/5}} + 4\frac{A_c^3(r)A_c^{'}(r)}{r^{18/5+n'}}\right] \tag{A8}$$

at $k \to 0$. Here, the next cases arise: $6/5 < n' < 7/5$ and $7/5 < n' < 12/5$.

In the first of them, $6/5 < n' < 7/5$, the coefficients at two main terms in $\tilde{c}_c(0) - \tilde{c}_c(k)$ when $k \to 0$, will be determined just by the asymptotic values $A_\infty$ and $A_\infty^{'}$. Then, in analogy with the derivation of (A4), we obtain from (A8):

$$\tilde{c}_c(0) - \tilde{c}_c(k) = 4\pi\left(a_{4c} + \frac{1}{4}\right)\left[\frac{25 A_\infty^4 I_{14/5}}{126} k^{9/5} + \frac{4 A_\infty^3 A_\infty^{'} I_{8/5+n'}}{(n'+8/5)(n'+3/5)} k^{n'+3/5}\right]. \tag{A9}$$

Note, that when finding (A9) we have reduced the integrals to the form (A5) with the help of the repeated integration by parts:

$$\int_0^\infty \frac{d\zeta}{\zeta^s}\left(1 - \frac{\sin \zeta}{\zeta}\right) = \frac{I_s}{s(s-1)}, \quad 1 < s < 3. \tag{A10}$$

Then, let us rewrite equation (A1) as follows:

$$\frac{1}{\tilde{h}_c(k)} = \rho_c^2[\tilde{c}_c(0) - \tilde{c}_c(k)], \tag{A11}$$

where we have omitted the terms inessential for further analysis. Substituting expressions (A4) and (A9) into (A11), we obtain with necessary accuracy the next equation at $k \to 0$:

$$\frac{1}{4\pi A_\infty I_{6/5}} \left( k^{9/5} - \frac{A'_\infty I_{n'}}{A_\infty I_{6/5}} k^{n'+3/5} \right)$$

$$= 4\pi \left( a_{4c} + \frac{1}{4} \right) \rho_c^2 \left[ \frac{25 A_\infty^4 I_{14/5}}{126} k^{9/5} + \frac{4 A_\infty^3 A'_\infty I_{8/5+n'}}{(n'+8/5)(n'+3/5)} k^{n'+3/5} \right], \quad \frac{6}{5} < n' < \frac{7}{5}. \quad (A12)$$

Equating the coefficients at $k^{9/5}$ in this equation, we find, first of all, the expression for the constant $A_\infty$:

$$A_\infty = \left[ \frac{63}{200\pi^2 \rho_c^2 I_{6/5} I_{14/5}(a_{4c}+1/4)} \right]^{1/5}. \quad (A13)$$

From this expression, taking account of (A6) the important condition on the admissible values of $a_{4c}$ follows:

$$a_{4c} > -\frac{1}{4}. \quad (A14)$$

Furthermore, equating the terms $\sim k^{n'+3/5}$ in equation (A12) and taking into account conditions (A6) and (A14), we find that the only permissible value of $A'_\infty$ is

$$A'_\infty = 0, \quad \frac{6}{5} < n' < \frac{7}{5}. \quad (A15)$$

Thus, *the terms with $6/5 < n' < 7/5$ obviously cannot be present in the asymptotic expansion of the "critical" total PCF $h_c(r)$.*

So, let us consider $\tilde{c}_c(0) - \tilde{c}_c(k)$ at $7/5 < n' < 12/5$. Now, for $k \to 0$ the integral from the term with the asymptotic behavior $\sim r^{-18/5-n'}$ in (A8) will already converge at the upper limit. Therefore, at small $k$, this integral will begin with an *analytical* term $\sim k^2$. In this situation, a coefficient at $k^2$ will be determined by the behavior of the integrand in the whole region of variation of $r$, but not only at large distances $r$. However it is important that already the next term in $\tilde{c}_c(0) - \tilde{c}_c(k)$ is *non-analytic*. To determine this term let us turn to formula (A8) at $7/5 < n' < 12/5$ and consider separately the expression

$$Q(k;n') \equiv q(n')k^2 + \int_0^\infty dr\, r^2 \left[ 1 - \frac{\sin(kr)}{kr} - \frac{k^2 r^2}{3!} \right] \frac{A_c^3(r) A_c'(r)}{r^{18/5+n'}} \quad (A16)$$

in which we have specially isolated the analytical term $\sim k^2$ with the coefficient

$$q(n') = \frac{1}{3!} \int_0^\infty dr\, r^4 \frac{A_c^3(r) A_c'(r)}{r^{18/5+n'}}. \quad (A17)$$

The remaining integral in (A16) at $7/5 < n' < 12/5$ will lead again to non-analytic term $\sim k^{n'+3/5}$ in the limit $k \to 0$. A coefficient at $k^{n'+3/5}$ will again be determined by the asymptotic values $A_\infty$ and $A'_\infty$ (cf. (A9)), and thus at small $k$ we obtain:

$$Q(k;n') = q(n')k^2 + A_\infty^3 A'_\infty k^{3/5+n'} \int_0^\infty \frac{d\zeta}{\zeta^{8/5+n'}} \left(1 - \frac{\sin\zeta}{\zeta} - \frac{\zeta^2}{6}\right). \tag{A18}$$

Taking the integral in (A18) by parts several times, we can reduce it to the form (A5) for the interval $7/5 < n' < 12/5$:

$$\int_0^\infty \frac{d\zeta}{\zeta^{8/5+n'}} \left(1 - \frac{\sin\zeta}{\zeta} - \frac{\zeta^2}{6}\right) = -\frac{I_{n'-2/5}}{(n'+8/5)(n'+3/5)(n'-2/5)(n'-7/5)}. \tag{A19}$$

Pay attention to the *negativity* of this integral. As the result, taking $n'$ from the interval $7/5 < n' < 12/5$ and keeping the main terms in the limit $k \to 0$ we obtain the following expression:

$$\tilde{c}_c(0) - \tilde{c}_c(k) = 4\pi \left(a_{4c} + \frac{1}{4}\right) \left[\frac{25 A_\infty^4 I_{14/5}}{126} k^{9/5} + 4q(n')k^2 \right.$$
$$\left. - \frac{4 A_\infty^3 A'_\infty I_{n'-2/5}}{(n'+8/5)(n'+3/5)(n'-2/5)(n'-7/5)} k^{n'+3/5}\right]. \tag{A20}$$

Now, substituting expressions (A4) and (A20) into equation (A11) and retaining the main terms at $k \to 0$ we come to the equation (cf. (A12)):

$$\frac{1}{4\pi A_\infty I_{6/5}} \left(k^{9/5} - \frac{A'_\infty I_{n'}}{A_\infty I_{6/5}} k^{n'+3/5}\right) = 4\pi \left(a_{4c} + \frac{1}{4}\right) \rho_c^2 \left[\frac{25 A_\infty^4 I_{14/5}}{126} k^{9/5} + 4q(n')k^2 \right.$$
$$\left. - \frac{4 A_\infty^3 A'_\infty I_{n'-2/5}}{(n'+8/5)(n'+3/5)(n'-2/5)(n'-7/5)} k^{n'+3/5}\right], \quad \frac{7}{5} < n' < \frac{12}{5}. \tag{A21}$$

From here we again obtain expression (A13). Besides, from (A21) the integral condition follows:

$$q(n') \equiv 0, \quad \frac{7}{5} < n' < \frac{12}{5}. \tag{A22}$$

This condition could play a useful role in the modeling of the functions $A_c(r)$ and $A'_c(r)$ at all $r$ (however, now this is not our task).

At last, equating the coefficients at the terms $\sim k^{n'+3/5}$ in (A21) and using (A13), one can see that now, in contrast to (A15), non-zero values for the constant $A'_\infty$ are, in principle, admissible. The necessary condition for the existence of the values $A'_\infty \neq 0$ is the fulfillment of the equality

$$\frac{I_{n'-2/5}}{I_{n'}} = \frac{25 I_{14/5}}{504 I_{6/5}} \left(n' + \frac{8}{5}\right)\left(n' + \frac{3}{5}\right)\left(n' - \frac{2}{5}\right)\left(n' - \frac{7}{5}\right), \quad \frac{7}{5} < n' < \frac{12}{5}. \tag{A23}$$

As not complicated numerical calculation shows, the transcendental equation (A23) in the pointed out interval of $n'$ really has unique solution which is given by Eq. (15).

## Appendix B. The calculation of critical exponent $n''$

To find the exponent $n''$ and the form of the constant $A_\infty''$ in the asymptotic expression for the "critical" total PCF (16) let us continue expansion (A2), introducing a new function $A_c''(r)$, in order to guarantee the correct behavior of $h_c(r)$ at all $r$:

$$h_c(r) = \frac{A_c(r)}{r^{6/5}} + \frac{A_c'(r)}{r^{n'}} + \frac{A_c''(r)}{r^{n''}}, \tag{B1}$$

where $A_c''(r \to \infty) \to A_\infty''$. Being based on (B1) and acting as when deriving expression (A4), we find the main contributions into $\tilde{h}_c(k)$ at $k \to 0$:

$$\tilde{h}_c(k) = 4\pi A_\infty I_{6/5} k^{-9/5} + 4\pi A_\infty' I_{n'} k^{n'-3} + 4\pi A_\infty'' I_{n''} k^{n''-3}, \tag{B2}$$

where the last term will be also divergent if $n'' < 3$. We shall make sure that the desired value of $n''$ will be less than $12/5$, so it will again be enough to restrict ourselves by the term $\sim h_c^4(r)$ in the "critical" direct PCF (8), based on approximation (B1). Assuming to use $[\tilde{h}_c(k)]^{-1}$ in (A11), we have from (B2) the next equation to within the main contributions at $k \to 0$:

$$\frac{1}{\tilde{h}_c(k)} = \frac{1}{4\pi A_\infty I_{6/5}} \left[ k^{9/5} - \frac{A_\infty' I_{n'}}{A_\infty I_{6/5}} k^{n'+3/5} - \frac{A_\infty'' I_{n''}}{A_\infty I_{6/5}} k^{n''+3/5} + \left(\frac{A_\infty' I_{n'}}{A_\infty I_{6/5}}\right)^2 k^{2n'-3/5} \right]. \tag{B3}$$

Further, to within the main terms consistent with (B1) and the asymptotic dependence (16), one can represent the "critical" direct PCF (8) in the form:

$$c_c(r) = \left(a_{4c} + \frac{1}{4}\right)\left\{\frac{A_c^4(r)}{r^{24/5}} + 4\frac{A_c^3(r)}{r^{18/5}}\left[\frac{A_c'(r)}{r^{n'}} + \frac{A_c''(r)}{r^{n''}}\right] + 6\frac{A_c^2(r)}{r^{12/5}}\left[\frac{A_c'(r)}{r^{n'}}\right]^2\right\}. \tag{B4}$$

Note, that the last term in the curly brackets of (B4) must be conserved because it has the same asymptotic behavior as the term $\sim r^{-18/5-n''}$ (see below). Now, the singular contributions into $\tilde{c}_c(0) - \tilde{c}_c(k)$ in the limit $k \to 0$ will be determined not only by the constants $A_\infty$ and $A_\infty'$ but by the constant $A_\infty''$ also. Using the explicit form (B4) and acting as when obtaining (A20), we find to within the main terms at $k \to 0$:

$$\tilde{c}_c(0) - \tilde{c}_c(k) = 4\pi\left(a_{4c} + \frac{1}{4}\right)\left[\frac{25 A_\infty^4 I_{14/5}}{126} k^{9/5} + 4q(n',n'')k^2\right.$$

$$-\frac{4 A_\infty^3 A_\infty' I_{n'-2/5} k^{n'+3/5}}{(n'+8/5)(n'+3/5)(n'-2/5)(n'-7/5)} - \frac{4 A_\infty^3 A_\infty'' I_{n''-2/5} k^{n''+3/5}}{(n''+8/5)(n''+3/5)(n''-2/5)(n''-7/5)}$$

$$\left. -\frac{6(A_\infty A_\infty')^2 I_{2n'-6/5} k^{2n'-3/5}}{(2n'+2/5)(2n'-3/5)(2n'-8/5)(2n'-13/5)}\right], \quad \frac{7}{5} < n' < n'' < \frac{12}{5}. \tag{B5}$$

Here $4q(n',n'')k^2$ (cf. (A20)) is an analytical contribution due to "non-dangerous" terms with the exponents $n'$ and $n''$ in (B4) (for our purposes, we will not need the explicit form of this contribution).

Substituting expressions (B3) and (B5) into (A11), we come to the equation:

$$\frac{1}{4\pi A_\infty I_{6/5}}\left[k^{9/5} - \frac{A_\infty' I_{n'}}{A_\infty I_{6/5}} k^{n'+3/5} - \frac{A_\infty'' I_{n''}}{A_\infty I_{6/5}} k^{n''+3/5} + \left(\frac{A_\infty' I_{n'}}{A_\infty I_{6/5}}\right)^2 k^{2n'-3/5}\right]$$

$$= 4\pi\left(a_{4c} + \frac{1}{4}\right)\rho_c^2\left[\frac{25 A_\infty^4 I_{14/5}}{126} k^{9/5} + 4q(n',n'')k^2\right.$$

$$-\frac{4 A_\infty^3 A_\infty' I_{n'-2/5} k^{n'+3/5}}{(n'+8/5)(n'+3/5)(n'-2/5)(n'-7/5)} - \frac{4 A_\infty^3 A_\infty'' I_{n''-2/5} k^{n''+3/5}}{(n''+8/5)(n''+3/5)(n''-2/5)(n''-7/5)}$$

$$\left. -\frac{3(A_\infty A_\infty')^2 I_{2n'-6/5} k^{2n'-3/5}}{8(n'+1/5)(n'-3/10)(n'-4/5)(n'-13/10)}\right], \quad \frac{7}{5} < n' < n'' < \frac{12}{5}. \tag{B6}$$

Now, one should equate the coefficients at the equal powers of $k$ in both sides of equation (B6). In this case, as well as earlier, equalities (A13) and (A23) will follow, but as an integral condition we obtain, instead of (A22):

$$q(n',n'') \equiv 0, \quad \frac{7}{5} < n' < n'' < \frac{12}{5}. \tag{B7}$$

However the situation with the rest of the terms in equation (B6) turns out to be rather curious. When trying to equate the terms $\sim k^{n''+3/5}$ in both sides of (B6) and assuming the existence of non-trivial solution for the constant $A_\infty''$, we immediately come to the equation for $n''$, exactly coinciding in form with equation (A23) for $n'$. This means that, taking into account the last term in (B1), nothing new is obtained, that is, this term should be eliminated, setting $A_\infty'' = 0$. Further, if we shall try to equate the terms $\sim k^{2n'-3/5}$ in (B6), then due to their certainly different signs, we should come again to the trivial solution (A15), but now for the interval $7/5 < n' < 12/5$, what is absurd in view of condition (15) for the existence of $A_\infty' \neq 0$. In reality, a non-trivial solution for $A_\infty''$ can be found non-contradictorily if we require the equality of the exponents

$$n'' + \frac{3}{5} = 2n' - \frac{3}{5}, \tag{B8}$$

which is equivalent to the equation (17) of the main text. Under the condition (B8), we must equate the resulting coefficients at the terms of the corresponding powers of $k$ in both sides of equation (B6). Then, taking into account (A13) and omitting intermediate calculations, we obtain the expression:

$$A_\infty^{''} = \frac{C_1}{C_2} \frac{(A_\infty^{'})^2}{A_\infty}, \qquad (B9)$$

where we have designated:

$$C_1 \equiv \left(\frac{I_{n'}}{I_{6/5}}\right)^2 + \frac{189 I_{2n'-8/5}}{100(n'+1/5)(n'-3/10)(n'-4/5)(n'-13/10)I_{14/5}}, \qquad (B10)$$

$$C_2 \equiv \frac{I_{2n'-6/5}}{I_{6/5}} - \frac{63 I_{2n'-8/5}}{50(n'+1/5)(n'-3/10)(n'-4/5)(n'-13/10)I_{14/5}}. \qquad (B11)$$

Using the explicit values of $n'$ from (15) and of the integrals (A5) we find $C_1 = 1.99953...$ and $C_2 = 1.42476...$, so that the ratio figuring in (B9) is $C_1/C_2 = 1.40341...$. Turn our attention to the fact that $A_\infty^{''} \geq 0$ (cf. (A6)); moreover, from (B9) it follows that $A_\infty^{''}$ can become zero only simultaneously with $A_\infty^{'}$. Note, at last, that the values of constants $A_\infty$ and $A_\infty^{'}$ can be set independently, whereas $A_\infty^{''}$ will be expressed through them by (B9). Thanks to this "genetic" connection between these constants one should consider the second and third terms in the asymptotic expansion (16) simultaneously.